# Combined effects of nonmetallic impurities and planned metallic dopants on grain boundary energy and strength

Zhifeng Huang [a, b], Fei Chen [a, *], Qiang Shen [a], Lianmeng Zhang [a], Timothy J. Rupert [b, c, *]

[a] State Key Lab of Advanced Technology for Materials Synthesis and Processing, Wuhan University of Technology, Wuhan 430070, China

[b] Department of Mechanical and Aerospace Engineering, University of California, Irvine, CA 92697, USA

[c] Department of Materials Science and Engineering, University of California, Irvine, CA 92697, USA

* Corresponding author: chenfei027@whut.edu.cn (F. Chen), trupert@uci.edu (T.J. Rupert)

**Abstract:**

Most research on nanocrystalline alloys has been focused on planned doping of metals with other metallic elements, but nonmetallic impurities are also prevalent in the real world. In this work, we report on the combined effects of metallic dopants and nonmetallic impurities on grain boundary energy and strength using first-principles calculations, with a Σ5 (310) grain boundary in Cu chosen as a model system. We find a clear correlation between the grain boundary energy and the change in excess free volume of doped grain boundaries. A combination of a larger substitutional dopant and an interstitial impurity can fill the excess free volume more efficiently and further reduce the grain boundary energy. We also find that the strengthening effects of dopants and impurities are dominated by the electronic interactions between the host Cu atoms and the two types of dopant elements. For example, the significant competing effects of metal dopants such as Zr, Nb, and Mo with impurities on the grain boundary strength are uncovered from the density of states of the *d* electrons. As a whole, this work deepens the field's understanding of the interaction between metallic dopants and nonmetallic impurities on grain boundary properties, providing a guide for improving the thermal stability of materials while avoiding embrittling effects.

## 1. Introduction

Grain boundaries play an important role in governing the mechanical, functional, and kinetic properties of a great many engineering materials, but these features are especially important for nanostructured materials [1-4]. Nanostructured materials exhibit many advantages compared to microcrystalline materials, including superior strength [5, 6] as well as increased resistance to wear [7] and fatigue [8]. However, one of the limitations of nanostructured materials is their lack of thermal stability, which is attributed to the high grain boundary fraction providing a large driving force for grain growth [9-11]. A number of experimental and theoretical research studies have shown that the thermal stability of nanostructured metals can be significantly improved by metallic dopant segregation at grain boundaries [12-17]. Murdoch and Schuh [15] built a grain boundary segregation enthalpy map for hundreds of binary alloys, which provides an important guide for both experimental and theoretical studies aimed at discovering thermally-stable nanocrystalline binary alloy combinations. Liu and Kirchheim [16] showed that the grain boundary energy, the key driving force for coarsening, can be reduced by metallic dopants to improve the stability of nanostructured materials. Most work on grain boundary segregation and stabilization has been focused on planned doping with metals [12-17], but such alloys will also likely contain common nonmetallic impurities incorporated during materials processing and service. For example, H, C, and O are often introduced by process control agents during mechanical alloying [17-20].    Nonmetallic impurities have also been found to play important role in grain size stabilization [20-25].    For example, He et al. [21] found that stress-driven grain boundary migration in nanocrystalline Al can be retarded by having an excess of O atoms at boundaries. Juárez et al. [20] showed that the dissolution of C has a positive effect on the thermal stability of an Fe–Zr nanocrystalline alloy.    However, while both planned metallic dopants and unplanned nonmetallic impurities will be present in the vast majority of nanostructured alloys, the combined effect of these dopants on thermal stability has not been studied in detail.

Metallic dopants have also been reported to play an important role in altering the mechanical strength of grain boundaries [14, 26-31].    For instance, a nanocrystalline Al–Mg alloy was reported to exhibit a yield strength much greater than the upper limit reported for traditional age-hardened Al alloys, with this high strength resulting from Mg segregating to the grain boundaries [28].    Wu et al. [29] found that the strengthening effect of metallic dopants on W grain boundaries depended on the type of grain boundary structure and the atomic radius of the added dopant.    At the same time, the effect of impurities on similar properties has been an active area of research [22, 32-36].    For example, the incorporation of a small quantity of nonmetallic H impurities can cause embrittlement [32-34].    In contrast, impurities such as B [35, 37-39] and C [40] have been reported to improve the grain boundary strength in some alloys.    Unfortunately, there are only a few studies in the literature that have focused on understanding how these two types of solutes combine to affect grain boundary strength, with this work often limited to a specific combination of added metal and impurity.    Yang et al. [41] found that N would eliminate the Mn-induced detrimental effect on the strength of an Fe grain boundary, while Zhong et al. [42] reported that P would enhance Mn-induced embrittlement in an Fe grain boundary. Zhang et al. also [43] found that Si could weaken the Na-induced embrittlement of Al grain boundary by forming strong Al–Si bonds at the interface. Unfortunately, these limited number of studies do not provide a comprehensive picture of the combined effects of these two types of solutes on strengthening or embrittling effects.

In this work, we report on the combined effect of a large variety of both metallic dopants (Al, Zn, Zr, Nb, Mo, Pd, Ag, and Bi) and nonmetallic impurities (H, B, C, N, O, Si, P, and S) on the grain boundary energy and strength of a Σ5 (310) grain boundary in Cu. We employ first-principles simulations to study this issue at the atomic and electronic levels [44]. First, we calculate the grain boundary energies and strengthening energies of interfaces with the metallic dopants, to provide a baseline for comparison when impurities are added. Next, we calculate the segregation energies of grain boundaries with both metallic dopants and nonmetallic impurities at various positions relative to one another. We find that the preferred sites of dopants and impurities are related to both their atomic radius and electronegativity. Grain boundary energy decreases as the atomic radius of both the dopants and impurities increases, because these atoms can more efficiently fill the excess free volume at the grain boundary. During our analysis of the mechanical effects, we divide the strengthening energy into mechanical and chemical contributions to provide a more nuanced picture of this effect. The mechanical contribution increases with increasing amounts of grain boundary expansion. However, for the majority of our samples, the main contribution of grain boundary strengthening or weakening comes from the chemical contribution, predominantly due to interactions between the $d$ states of the dopants and the host Cu atoms. For example, the $d$-states interactions are weakened when the $p$ and $s$ states of impurities create hybrid orbitals with the *d* states of the dopants, leading to competition between the two types of elements. As a whole, this work deepens the field's understanding of the combined effects of metallic dopants and nonmetallic impurities, which can provide a guide for tailoring the stability of a microstructure while avoiding embrittlement. While this work is motivated by the need for better nanostructured materials, alterations to the grain boundary energy and strength will also be important for coarse-grained materials.

## 2. Computational methods

Fig. 1(a) shows the model of a Σ5 (310) grain boundary in Cu. This type of grain boundary was chosen because there are four substitutional sites (marked as 1, 2, 3, and 4) and three interstitial sites (the pentagonal bipyramid (PBP), bitetrahedron (BTE), and cap trigonal prism (CTP) sites at the interface [32, 35]), making it an appropriate model for a systematic investigation. Prior studies have shown that the substitutional sites are the preferred sites for the metallic dopants, while the interstitial sites are the preferred sites for nonmetal impurities [30, 35]. The grain boundary specimen has dimensions of $7.267 \times 11.490 \times 28.086$ Å$^3$ and contains 112 atoms. Hereafter, we will refer to metallic atom additions as "dopants" and nonmetallic atoms as "impurities." Since the atomic radius and electronegativity of dopants and impurities might be key factors for alteration of grain boundary energy and strength [29, 35, 45] and a range of these parameters is sought, we choose Al, Zn, Zr, Nb, Mo, Pd, Ag, and Bi as the metallic dopants while selecting H, B, C, N, O, Si, P, and S as the nonmetallic impurities. The atomic radii [46] and electronegativities [47] of these choices are listed in Table 1, along with the values for Cu. For the grain boundary model with both a dopant and an impurity, the added atoms could have different relative spatial positions. We define the "Near" configuration as being when the dopant and the impurity bond with each other in the same periodic unit, while the "Far" configuration occurs when the dopant and the impurity stay as far as possible within the interface, occupying the sites locating in different periodic units. These two possibilities are illustrated in Fig. 1(b). First-principles calculations were performed with the Vienna ab-initio simulation package (VASP) using the projector augmented wave approach [48] and the Perdew-Burke-Ernzerhof exchange-correlation generalized gradient approximation (GGA) functional [49]. A plane-wave cutoff energy of 350 eV, k-point meshes of $3 \times 2 \times 1$, convergence energy of $10^{-5}$ eV/atom, and convergence atomic force of 0.01 eV/Å were used for all calculations to balance the accuracy and efficiency of calculations [35]. The atoms in grain boundary model were fully relaxed during the process of structural optimizations, except that the *z* coordinate of atoms on the outermost layers were fixed.

The propensity of a dopant X or an impurity Y to segregate to the grain boundary can be characterized by the segregation energy, $E_{seg}^{GB}$ [35, 50]:

$$E_{seg}^{GB[m,n]} = (E_{tot}^{GB[m,n]} - E_{tot}^{GB[0,0]}) - (E_{tot}^{bulk[m,n]} - E_{tot}^{bulk[0,0]}) \qquad (1)$$

where $E_{tot}^{GB}$ and $E_{tot}^{bulk}$ are the total energies of the grain boundary and bulk models. The [*m*, *n*] represents the model contains *m* dopant atoms and *n* impurity atoms, thus the [0, 0] indicates the pure Cu model without dopant or impurity. The bulk model has the exact same dimension and number of atoms as the grain boundary model, but no grain boundary [35]. Near and Far configurations are also considered in the bulk models with both a dopant X and an impurity Y. $E_{tot}^{bulk[m,n]}$ is the total energy of the lowest energy bulk model. A more negative segregation energy indicates the model is more energetically stable [35, 50]. The grain boundary energy, $\gamma$, with X or Y can be calculated as [26, 35, 51]:

$$\gamma[m,n] = \frac{E_{tot}^{GB[m,n]} - E_{tot}^{bulk[m,n]}}{S} \qquad (2)$$

where $S$ is the cross-sectional area of the simulation cell (i.e., the grain boundary area). Eq. (1) and (2) contain some of the same terms and are related, with a dopant X or an impurity Y with the strong ability to segregate to the grain boundary usually significantly reducing the grain boundary energy.

The strength of the grain boundary can be represented by the separation energy, $E_{sep}$, which is defined as the energy needed to separate the grain boundary into two free surfaces (the lower surface and the upper surface) [52]:

$$E_{sep} = E_{tot}^{FS\_lower} + E_{tot}^{FS\_upper} - E_{tot}^{GB} \qquad (3)$$

where $E_{FS\_lower}^{tot}$ and $E_{FS\_upper}^{tot}$ are the total energy of the lower surface and the upper surface, respectively. For the grain boundary with just one dopant or impurity, we assign that the dopant or impurity stays on the lower surface, meaning the upper surface has the same atomic configuration as the clean Cu free surface. For the grain boundary with both a dopant and an impurity, there are 48 relative positions for dopant-impurity pairs with the four substitutional sites of the dopant, the three interstitial sites of the impurity, the Near and Far configurations, and the lower and upper surface. To allow for a reasonable number of calculations, we assume that the dopant and the impurity stay on the lower surface after the separation. The free surface model was obtained using the exact same dimensions as the grain boundary model but with the upper half removed, as shown in Fig. 1(c). Similar models and assumptions have also been used in previous work in the Σ5 (310) grain boundary in Cu with multi substitutional and interstitial solutes [37, 52]. The strengthening energy, $E_{str}$, can be defined as the difference between the separation energies of the clean grain boundary and the grain boundary with X or Y [35, 50]:

$$E_{str}^{[m,n]} = E_{sep}^{GB[0,0]} - E_{sep}^{GB[m,n]} \qquad (4)$$

Calculations of the free surface energies are important for these grain boundary strength calculations, so there are two important scenarios to consider. One measure of the grain boundary strength would have the dopants and impurities on the free surface stay at the lowest energy sites calculated for the grain boundary, representing a fast fracture process where no diffusion along the crack surface is allowed (denoted as "Fast"). Alternatively, one would allow the dopants and impurities to find the lowest energy sites along the free surface of the crack, which represents a slow fracture case where diffusion can and does occur (denoted as "Slow"). Both calculations were carried out, providing upper and lower bounds for the strengthening effect. A negative value of the strengthening energy means that the dopant or the impurity will enhance the grain boundary strength, while a positive value suggests a detrimental effect on strength. In both the segregation energy and the strengthening energy, negative energies would be preferred to achieve a more stable grain structure and to strengthen the boundary against cracking.

Furthermore, the strengthening energy can be divided into mechanical ($E_{\text{mech}}$) and chemical ($E_{\text{chem}}$) contributions, which reflect the effects resulting from the structural distortion of the boundary and the electronic interaction between the host Cu and solutes, respectively [29, 30, 37, 52]. Multiple essential scenarios are defined and shown in Fig. 2. System A is the relaxed clean grain boundary and clean free surface, with only Cu atoms. System B is the grain boundary and free surface with a Cu vacancy, which is generated by removing a Cu atom from the corresponding models in the system A but with no further structural relaxation. The site for the Cu vacancy is the same site as the site of a substitutional dopant in system D. System C is the grain boundary and free surface with the structural distortion caused by adding the substitutional dopant or the interstitial impurity, which is generated by removing the substitutional dopant or the interstitial impurity in system D but no further relaxation. System D is the final relaxed grain boundary and free surface with a substitutional dopant or an interstitial impurity. The path to introduce a substitutional dopant to the grain boundary can be thought of as removal of a Cu atom at the grain boundary (A → B), adjustment of the space at the grain boundary (B → C), and then finally addition of the substitutional dopant at the grain boundary (C → D). The path for an interstitial impurity is A → C → D without removing a Cu at the grain boundary. The path for adding both a substitutional dopant and an interstitial impurity is A → B → C → D. The mechanical contribution to the strengthening energy will be the difference between the separation energies in the systems B and C for a grain boundary with a metallic dopant and a grain boundary with both a dopant plus an impurity:

$$E_{mech} = E_{sep}^{(B)} - E_{sep}^{(C)} \tag{5}$$

Since there is no removal of a host Cu atom, the mechanical contribution for the grain boundary with an interstitial impurity is:

$$E_{mech} = E_{sep}^{(A)} - E_{sep}^{(C)} \tag{6}$$

The chemical contribution is then the total strengthening effect with the mechanical contribution subtracted:

$$E_{chem} = E_{str} - E_{mech} \tag{7}$$

## 3. Results and discussion

### 3.1 Grain boundaries with only metallic dopants

The segregation energies of different metallic dopants at the grain boundary are shown in Fig. 3, which demonstrates that the site 1 is the preferred site for all of the dopants. Fig. 4(a) then shows the calculated grain boundary energies with dopants at the site 1. All of the grain boundary energies are lower than the 880.37 $mJ/m^2$ value of the clean grain boundary [35], meaning that all dopants studied here can increase the grain boundary stability. Fig. 4(b) shows that the grain boundary energy decreases as the atomic radius of the dopant increases. In short, dopants with larger atomic radii have an increased ability to segregate to the grain boundary and stabilize it, which is consistent with prior findings in the literature [27, 53].

The calculated strengthening energies of grain boundaries with dopants at site 1 under the Fast and Slow fracture processes are shown in Fig. 4(c). We can see that strengthening energies under Fast fracture are always lower than those under Slow fracture. This is expected, since the energy of the boundary will be the same in these two cases but the energy of the free surface with dopant is always lower if Slow fracture occurs (a lower energy free surface state is found by way). In the end, the effect of each dopant on the grain boundary strength are similar under Fast and Slow fracture, with only the absolute value altered. Addition of Zr, Nb, or Mo will significantly improve the grain boundary strength, while Pd, Ag, and Al increase the strength to a small degree. In contrast, Zn and Bi weaken or embrittle the grain boundary. Our simulated results are in good agreement with the experimental observations that Zr [6] and Nb [54] can strengthen while Bi [55] embrittles nanocrystalline Cu. To avoid overcomplicating the discussion of our results, for the remainder of this paper we only discuss the strengthening effects under the Fast fracture condition. Fig. 4(d) presents the total strengthening energy into the host with the values broken down into mechanical and chemical contributions [37, 52]. The values of mechanical contributions of all dopants are positive, signifying weakening/embrittlement. The values of chemical contributions are negative, with the lone exception being Bi, indicating that the chemical interactions between the host Cu and the dopants usually improve the strength of the boundary. As a whole, the trend for the chemical contribution is similar to the trend for the total strengthening energies, indicating that the chemical contribution dictates the overall grain boundary strengthening behavior.

To help understand the importance of chemical effects, the density of states [38, 39, 56] for dopants in site 1 and the closest Cu atom in site 2 are calculated to investigate the electronic interactions (Fig. 5). The density of states data for the Cu atom appear as thin curves while the dopant data appears as thick curves. For the main-group metallic dopants (Al and Bi), the electronic interactions between the dopants and Cu are mainly reflected in the hybridizations between the $p$ states of the dopants and the $d$ states of Cu. For the transition metallic dopants except Zn, the main electronic interactions occur on the hybridization between the $d$ states of both dopants and Cu. Furthermore, for the transition metallic dopants with the $d$ orbitals fully occupied (Zn, Pd, and Ag), the $d$ states of dopants are very localized and form a sharp peak. The lack of overlap between the sharp $d$ states peak of Zn and the $d$ states peak of Cu indicates almost no interaction between Zn and Cu atoms in this system. The sharp peaks of Ag localize at the edge of the $d$ states peak of Cu, which means only a small fraction of the electrons in the $d$ states of Ag interact with Cu. The peaks of Pd are also sharply localized but in this case the peaks overlap with the *d* states of Cu, indicating a stronger interaction of Pd with Cu and explaining why Pd had the strongest strengthening effect of the transition metal dopants with the *d* orbitals fully occupied. In contrast, for the transition metallic dopants without the *d* orbitals fully occupied (Zr, Nb, and Mo), the *d* states of the dopants are more evenly distributed in the range of the *d* states of Cu, suggesting strong electronic interaction between the dopants and Cu. Referring back to Fig. 4(d), the strengthening effect of metallic dopants is closely correlated to the electronic interaction between the dopants and Cu atoms. The strengthening effect of dopants without *d* orbitals fully occupied is the highest, due to stronger interactions between the *d* states of the dopants and Cu. In addition, with the exception of Zn, it is clear that the degree of the overlap between the *d* states of the transition metallic dopants and Cu is larger than the overlap between the *p* states of the main-group metallic dopants and the *d* states of Cu atom. This indicates a stronger electronic interaction between the transition metallic dopants and Cu compared to the main-group metallic dopants. The strengths of grain boundaries with Zr, Nb, and Mo are significantly better than Pd, Ag, and Al, with boundaries doped with Zn and Bi being the worst.

## 3.2 Zr-segregated grain boundary with nonmetallic impurities

Since there are four substitutional sites for metallic dopants and three interstitial sites for nonmetallic impurities as shown in Fig 1, there are 24 relative positions between a given set of dopant plus impurity when considering the Near and Far configurations. In this work, eight dopants (Al, Zn, Zr, Nb, Mo, Pd, Ag, and Bi) and eight impurities (H, B, C, N, O, Si, P, and S) are considered, which means that 1536 grain boundary models would be needed to calculate all possible relative sites between the metallic dopants and nonmetallic impurities. In addition, a huge number of bulk models and free surface models would also need to be considered to calculate the segregation, grain boundary, and strengthening energies. Therefore, to make the problem more tractable, we temporarily focus on segregation energies in models with the metallic dopant Zr and common impurities C and O at different relative positions, with the results shown in Figs. 6(a) and (b). Zr by itself significantly reduces the grain boundary energy and improves the grain boundary strength, as shown in Fig. 4, so it is a good choice to provide a baseline. The segregation energies are relative lower when Zr occupies site 1, regardless of where the C or O impurity is located, which suggests that nonmetallic impurities do not determine the preferred site of metallic dopants. In Figs. 6(a) and (b), the lowest energy models contain C and O at the PBP site that is close to Zr (i.e., in the Near configuration). However, since the atomic radius of other impurities such as Si, P, and S are much larger, it is not possible to say that the Near configuration is always preferred. To this end, the segregation energies of the grain boundaries with Zr fixed at site 1 and impurities at different interstitial sites were calculated, taking care to consider both the Near and Far configurations (Fig. 6(c)). On this figure, if a data point does not appear, it means that the impurity moved from the chosen site when the system was relaxed, which means that the site in that case was not a possibility. H, C, N, and O prefer the PBP site, while B, Si, P, and S choose to occupy the CTP site. The relative smaller impurities H, B, C, N, and O prefer to bond with Zr in the Near configuration, while the larger impurities Si, P, and S prefer to stay far away from Zr in the Far arrangement.

The lowest energy grain boundary models with Zr and various impurities were then used to calculate the grain boundary energy. The relationship between the grain boundary energy and the atomic radius of nonmetallic impurities is plotted in Fig. 7(a). It is obvious that the grain boundary energy of the Zr-plus-impurity grain boundaries decreases as the atomic radius of the nonmetallic impurities increases. All the grain boundary energies in this figure are below the 568.75 mJ/m$^2$ grain boundary energy of the Zr-segregated interface, which is above the scale of the figure, meaning that adding impurities can further reduce the grain boundary energy. The strengthening energies of the grain boundary with Zr and impurities are shown in Fig. 7(b), with the data plotted as a function of the electronegativity of the impurities. B can further improve the strength of a Zr-segregated grain boundary while the other impurities weaken the interface. The strengthening energy of a Zr-segregated grain boundary increases with increasing electronegativity of the impurities from a given period in the periodic table (the two lines show which elements are in the same period). Furthermore, the strengthening energies are divided into mechanical and chemical contributions, as shown in Fig. 7(c). The results show that the strengthening energy is again mainly dependent on the chemical effects, which means the electronic interactions between the host Cu atoms, Zr dopant, and the impurities determine the grain boundary strength. The details of these electronic interactions will be discussed in more detail in Section 3.4.

## 3.3 Synergistic reduction of grain boundary energy by dopants and impurities

Next, we investigate the energetics of grain boundaries with a wider variety of dopants and impurities, no longer restricting our scope to Zr. The dopants studied are Al, Zn, Zr, Nb, Mo, Pd, Ag, and Bi, while the impurities are B, C, O, and Si. B, C, O and Si are chosen because they differ greatly in the atomic radius and prefer different sites at the grain boundary. Since Fig. 6 showed that site preference does not change when combining dopants and impurities, we maintained this feature to allow for efficient computation. Thus, dopants take site 1 in the boundary, while B and Si occupy the CTP site and C and O occupy the PBP site. The segregation energies of grain boundaries with both dopants and impurities at the Near and Far configurations are shown in Fig. 8. The small impurities B, C, and O prefer to stay close to the dopants such as Al, Zn, Zr, Nb, and Mo, but prefer to be further away from dopants such as Pd, Ag, and Bi. The large impurity Si prefers to remain far away from all of the metallic dopants. The variation observed for the smaller impurities can be explained by combining the information about atomic radius and electronegativity. Electronegativity is generally used to evaluate the ability of an atom to attract electrons towards itself [47, 57]. Therefore, one can hypothesize that an impurity will prefer to bond with a dopant when the electronegativity difference between the impurity and the dopant is bigger than the difference between the impurity and Cu. Table 1 shows that this is in fact true for our calculations. The dopants with electronegativity values lower than the value for Cu (1.8), such as Al (1.5), Zn (1.5), Zr (1.5), Nb (1.7), and Mo (1.6), prefer to bond with impurities. In contrast, Pd (2.0), with a higher electronegativity than Cu, prefers to stay far away from the impurities. Ag and Bi have similar electronegativity values as Cu, but the atomic radii of Ag (1.339 Å) and Bi (1.520 Å) are larger than Cu (1.173 Å) [46]. In this case, without an electronegativity to drive the preference, B, C, and O prefer to stay far away from Ag and Bi simply because there is more room in the interstitial site surrounded by only Cu atoms. Similarly, the atomic radius of Si (1.173 Å) is by far the largest of the impurities, meaning it will have the most trouble fitting into the interstitial sites due to this size. As a result, Si prefers to remain far away from the metallic dopants since they are all larger than Cu and a Near configuration would result in large structural distortions.

The relationship between the grain boundary energy and the atomic radius of each metallic dopant is plotted in Fig. 9(a), where the grain boundary energies are calculated from the lowest energy grain boundary models. It is obvious that the grain boundary energy decreases with increasing the atomic radius of dopants, as shown previously in Fig. 4. However, the addition of an impurity appears to shift this curve downwards by an amount that is element-dependent. For example, the grain boundary energies of a sample with dopants plus the large impurity Si are significantly lower than that of grain boundaries with dopants plus the smaller impurities B, C, and O. Prior work has shown that the grain boundary energy increases with increasing excess free volume of grain boundaries in face centered cubic metals [58-60]. Therefore, we hypothesize that the reduction in grain boundary energy, which is greatest for large dopants and large impurities, occurs because this free volume is being filled and reduced. The grain boundary excess free volume $\Omega[m,n]$ can be calculated as [58, 59]:

$$\Omega[m,n] = \frac{V_{GB[m,n]} - V_{bulk[m,n]}}{S} \tag{8}$$

where $V_{GB}$ and $V_{bulk}$ are the total volumes of the grain boundary region and the corresponding bulk Cu, respectively. The bulk model has the same cross-sectional area, $S$, and number of Cu atoms as the grain boundary model. The relationship of the change of grain boundary energy ($\gamma[m,n] - \gamma[0,0]$) and the change of the excess free volume ($\Omega[m,n] - \Omega[0,0]$) is plotted in Fig. 9(b), where grain boundaries with single dopants, single impurities, and both dopants and impurities are included. A general trend is found where the grain boundary energy decreases as the excess free volume decreasing, which confirms our hypothesis that reduction of excess free volume is a key component of reducing the grain boundary energy.  Herein, a combination of a larger metallic dopant and a larger interstitial impurity can further reduce the grain boundary energy since they can more efficiently fill the excess free volume of grain boundary.  For example, dopants plus the largest nonmetallic impurity Si have the greatest synergistic effect to reduce the grain boundary energy.

Previous works on Cu have also shown that the grain boundary energy decreases when increasing the interfacial coverage of dopants [27, 53], which supports the scientific concept presented here.  In addition, a number of literature reports have shown that grain boundary energy increases with increasing excess free volume by comparing distinct grain boundaries in various face centered cubic metals, such as Cu [60, 61], Ni [60, 62], Al [58, 59, 62], and Au [61].  Considering the similar atomic structures of grain boundary in face centered cubic metals, one can predict that the nanocrystalline structural stability of these metals will be improved by introducing large dopants and interstitial impurities to fill the excess free volume of grain boundaries.

### 3.4 Competition and synergy between dopants and impurities concerning strength

Fig. 10 shows the strengthening energies of grain boundaries with both dopants and impurities.  A wide variety of behaviors are observed, with both overall strengthening and weakening being found.  For example, B addition shifts the curves downwards, having a positive effect on grain boundary strength, even if it cannot overcome the negative effect of the metal dopant in the case of Bi.  In contrast, O always has the worst embrittling effect and always pushes the overall grain boundary strengthening effect into the positive values on Fig. 10, suggesting that O incorporated during materials processing will have a large negative effect on mechanical properties. The strengthening energies are again divided into mechanical and chemical contributions in Fig. 11. The values of the mechanical contributions are positive/weakening, except for a few very small negative/strengthening values for B, indicating that the mechanical contribution of impurities is generally to embrittle the boundary. As shown in Fig. 2, the mechanical contribution mainly originates from the local structural expansion at the interface [52]. The grain boundary expansion, $\Delta V$, can be defined as:

$$\Delta V = V_{(C)} - V_{(B)} \qquad (9)$$

where $V_{(C)}$ and $V_{(B)}$ are the total volume of the grain boundary regions in system C and B in Fig. 2, respectively. Fig. 12 shows the relationship between the mechanical contribution and the grain boundary expansion, where it is clear that the mechanical contribution becomes more positive/embrittling as the grain boundary expansion becomes larger. As was also observed for the samples with dopants only, the chemical contribution dominates the overall strengthening effect and the trends in the total effect tend to mimic the changes in the chemical contribution. To understand these trends, we focus on discussing the electronic interactions between dopants and impurities in the following section, with an eye for uncovering combined effects between the two solute species.

The combined effects, which can be either synergistic or competing, of metallic dopants and nonmetallic impurities on grain boundary strength are of great interest, in order to fulfill our original goal of providing a guide for finding potentially useful combinations of dopants and impurities to guide materials processing. The combined effect ($E_{str}^{comb}$) can be studied by taking the difference between the strengthening energy of the grain boundary with both dopants and impurities ($E_{str}^{GB[m,n]}$) and the sum of the strengthening energies of the grain boundary with single dopants ($E_{str}^{GB[m,0]}$) and the grain boundary with single impurities ($E_{str}^{GB[0,n]}$):

$$E_{str}^{comb} = E_{str}^{GB[m,n]} - E_{str}^{GB[m,0]} - E_{str}^{GB[0,n]} \qquad (10)$$

A negative value of $E_{str}^{comb}$ would signal that there is a synergistic effect of dopants and impurities, with the combined effect being stronger than the sum of its parts. In contrast, a positive value would signal that the dopant and impurity compete, giving an effect that is weaker than the sum of its parts. Finally, a value near zero will indicates that combined effect is very weak, meaning the total effect is simply the sum of the two contributions and there is no meaningful interaction between the two species that alters the boundary properties.

First, we investigate the combined effects of Zr and impurities on the grain boundary strength in Fig. 13(a). For all of the combinations, there is a competing effect between Zr and impurities, with this effect becoming worse as the atomic number of the impurities from a given period in the periodic table increases. Fig. 13(b)-(e) show the combined effects of all of the different dopants and the impurities B, C, O, and Si. For dopants and impurities where the Far configuration is preferred, the combined effect is near zero, indicating very little interaction. For dopants and impurities that are most stable in the Near configuration, only the dopants Al and Zn give values close to zero. In contrast, the combined effect of the dopants Zr, Nb, and Mo with impurities result in significantly positive values, indicating a strong competing effect.

Since the density of states data between the Nb/Mo and Cu are very similar to that of Zr and Cu, as shown in Fig. 5, we hypothesize that the origin of the competing effects between these dopants and impurities may be similar as well. To understand the reasons behind competing effects on grain boundary strength, we explore the details of the electronic interactions for the samples with Zr and various impurities. The density of states data for the Zr atom, the impurity atom (Y), and the closest Cu atom in site 2 are shown in Fig. 14. The density of states for Cu appear as thin curves, Zr appear as heavy curves, and the impurity Y appear as dashed curves. For the grain boundary with Zr and H, the $s$ states of H interact with the $d$ states of Zr but not with the $d$ states of Cu, suggesting that H has a stronger ability to bond with Zr. As a result, the electronic interaction between the $d$ states of Zr and Cu is weakened, which can be the cause behind the large embrittling chemical contribution shown in Fig. 13(a). For the grain boundaries with Zr and the second period impurities B, C, N, and O, the main electronic interactions occur in the $p$ states of impurities and the $d$ states of Zr and Cu. The $p$ states of impurities interact much more strongly with the $d$ states of Zr than the $d$ states of Cu, again indicating that Zr has the stronger ability to bond with these impurities. In addition, the main peaks of the $p$ states for the impurities move to the lower-energy regions as the atomic number of the impurities increase, further weakening the interaction with the $d$ states of Cu. At the same time, the overlap between the $p$ states of the impurities and the $d$ states of Zr are becoming stronger as the atomic number increases. In total, the interaction between the $d$ states of Cu and Zr are weakened by the strong electronic interactions between Zr and impurities. For the grain boundaries with Zr and the third periodic impurities Si, P, and S, although Zr stays away from these impurities in the Far configuration, the electronic interaction still occurs between the $s$ states of impurities and the $d$ states of Zr and Cu. As the atomic number of these impurities increases, the main peaks of the $s$ states of impurities move to the lower-energy regions and overlap more with the $d$ states of Zr. As a result, the electronic interactions between the *d* states of Zr and Cu are weakened by the interactions between Zr and these impurities, again giving a weakening effect that increases with increasing atomic number. Looking back to Fig. 13, it is clear that the competing effect between the dopants such as Zr, Nb, and Mo and the impurities becomes more pronounced with increasing atomic number of the impurities within a given period.

The results above, particularly Figures 9 and 10, can provide a guide for improving the stability of nanostructured materials while avoiding deleterious embrittlement.  For example, introducing transition metals without fully occupied *d* orbitals (such as Zr, Nb and Mo) into nanocrystalline Cu can significantly improve the thermal stability while also reducing the tendency for grain boundary fracture.  These stabilizing and strengthening effects can be further improved when combined with the nonmetallic impurity B.  Si may also be an acceptable impurity since the grain boundary energy can be dramatically reduced while the strength is only slightly decreased.  As mentioned in the introduction, C and O are common elements in process control agents for mechanical alloying and are often incorporated into Cu-based nanocrystalline alloys [17-20].  Although both improve the grain boundary stability, O has a large detrimental effect on the grain boundary strength.  Therefore, avoiding O contamination is essential in the preparation process for Cu-based nanostructured alloys.  In contrast, C only has a small negative effect on strength and may be acceptable due to the compromise that is often required for real-world materials processing.  It is worth reiterating that the strengthening effect is determined by the chemical interactions in this work, meaning that the grain boundary strengthening trends reported here may be different for different base metals.  For example, prior experimental work showed that O can improve the grain boundary strength in nanocrystalline Al [22]. The embrittling and strengthening effects of O in Cu and Al, respectively, originate from the different chemical interactions. Moreover, these chemical effects will be sensitive to the electron band structure of the base metal. For example, the d-band is fully filled in Cu. In contrast, W is a common structural metal with a very different electronic band structure, so the chemical interactions will likely be different and should be investigated if W-rich nanocrystalline alloys are of interest (see, e.g., [29]). In summary, for the stability of nanocrystalline alloys, the finding that both large metallic dopants and nonmetallic impurities would be useful additives can be generalized to other face centered cubic metals. When considering the mechanical properties of possible nanocrystalline alloys, one would need to pay additional attention to the chemical interactions between the solvent and solute, as these dominate in the Cu-rich alloys studied here.

## 4   Summary and conclusions

In conclusion, the combined effects of metallic dopants (Al, Zn, Zr, Nb, Mo, Pd, Ag, and Bi) and common nonmetallic impurities (H, B, C, N, O, Si, P, and S) on grain boundary energy and strength of a $\Sigma 5$ (310) grain boundary is Cu was investigated using first-principles calculations. The following specific conclusions can be drawn:

- The relative spatial positions of dopants and impurities are related to the atomic radii and electronegativity values. The dopants with less electronegativity than Cu prefer to bond with the impurities, as long as the atomic radii of the dopants and impurities are not extremely large.

- The grain boundary energy decreases as the excess free volume of the grain boundary

decreases. Therefore, a combination of a larger substitutional dopant and a larger interstitial impurity can more efficiently fill the free volume to further reduce the grain boundary energy.

- The strengthening/weakening effects of dopants and impurities mainly originate from the electronic interactions with Cu. Zr, Nb, and Mo can significantly enhance the grain boundary strength because of the strong interactions between the $d$ states of the dopants and Cu. In contrast, the strong interactions between the $s/p$ states of the impurities and the $d$ states of the dopants will dramatically reduce any strengthening effect.

As a whole, this work deepens the understanding of the combined effects of metallic dopants and nonmetallic impurities from the atomic and electronic levels, which can provide a guide on improving the stability and avoiding embrittlement of nanostructured materials. Such considerations are extremely important, as real-world nanocrystalline alloys typically contain a combination of planned or intentional dopants and unplanned or unintentional impurities.

**Acknowledgement:**

This research was supported by the U.S. Army Research Office under Grant W911NF-16-1-0369, the National Natural Science Foundation of China (51472188, 51521001, and 51872217), Natural Research Funds of Hubei Province (No. 2016CFB583), the Fundamental Research Funds for the Central Universities in China and the "111" project (B13035), and the National Key Research and Development Program of China (No. 2017YFB0310400 and

2018YFB0905600).

Table 1. The atomic radius (Å) [46] and electronegativity [47] of selected dopants and impurities, along with the values for Cu.

| **Metallic dopant** | **Al** | **Zn** | **Zr** | **Nb** | **Mo** | **Pd** | **Ag** | **Bi** | **Cu** |
|---|---|---|---|---|---|---|---|---|---|
| Atomic radius | 1.248 | 1.249 | 1.454 | 1.342 | 1.291 | 1.278 | 1.339 | 1.520 | 1.173 |
| Electronegativity | 1.5 | 1.5 | 1.5 | 1.7 | 1.6 | 2.0 | 1.8 | 1.8 | 1.8 |
| **Nonmetallic impurity** | **H** | **B** | **C** | **N** | **O** | **Si** | **P** | **S** | |
| Atomic radius | 0.320 | 0.800 | 0.771 | 0.700 | 0.660 | 1.173 | 1.100 | 1.040 | |
| Electronegativity | 2.1 | 2.0 | 2.5 | 3.0 | 3.5 | 1.8 | 2.1 | 2.5 | |

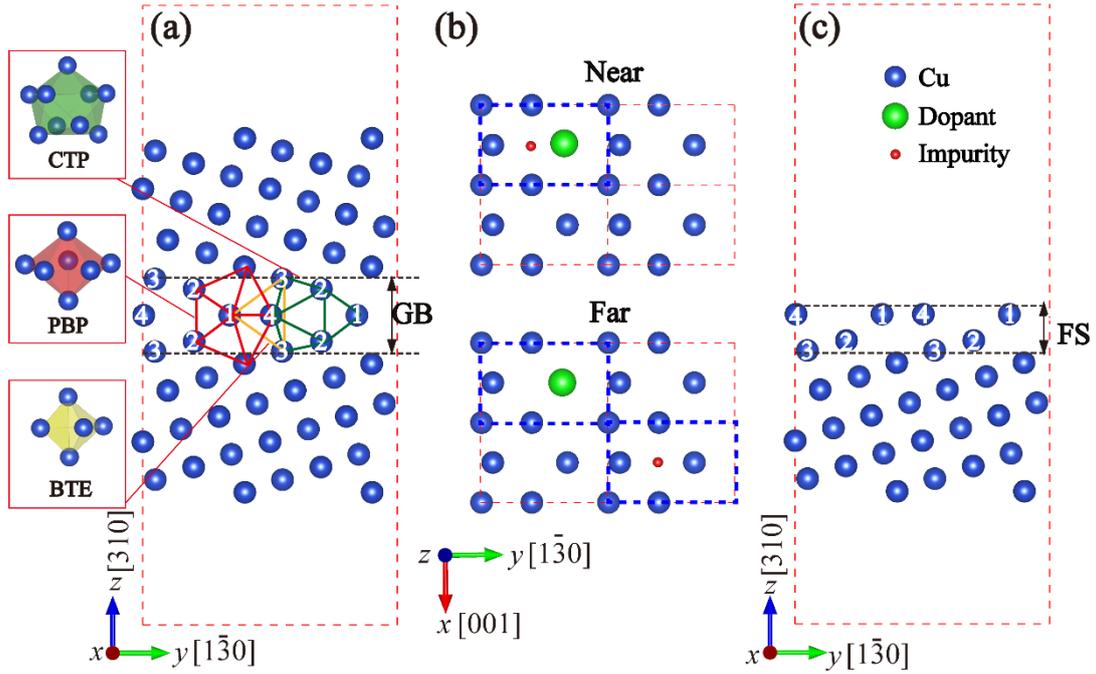

Fig. 1. Schematic illustrations of (a) Cu Σ5 (310) grain boundary (GB), (b) the Near and Far configurations of the grain boundary with both a substitutional dopant and an interstitial impurity, and (c) Cu Σ5 (310) free surface (FS). For the Near configuration, the dopant and impurity bond with each other and stay in the same periodic unit, while for the Far model, the dopant and impurity stay in different periodic units.

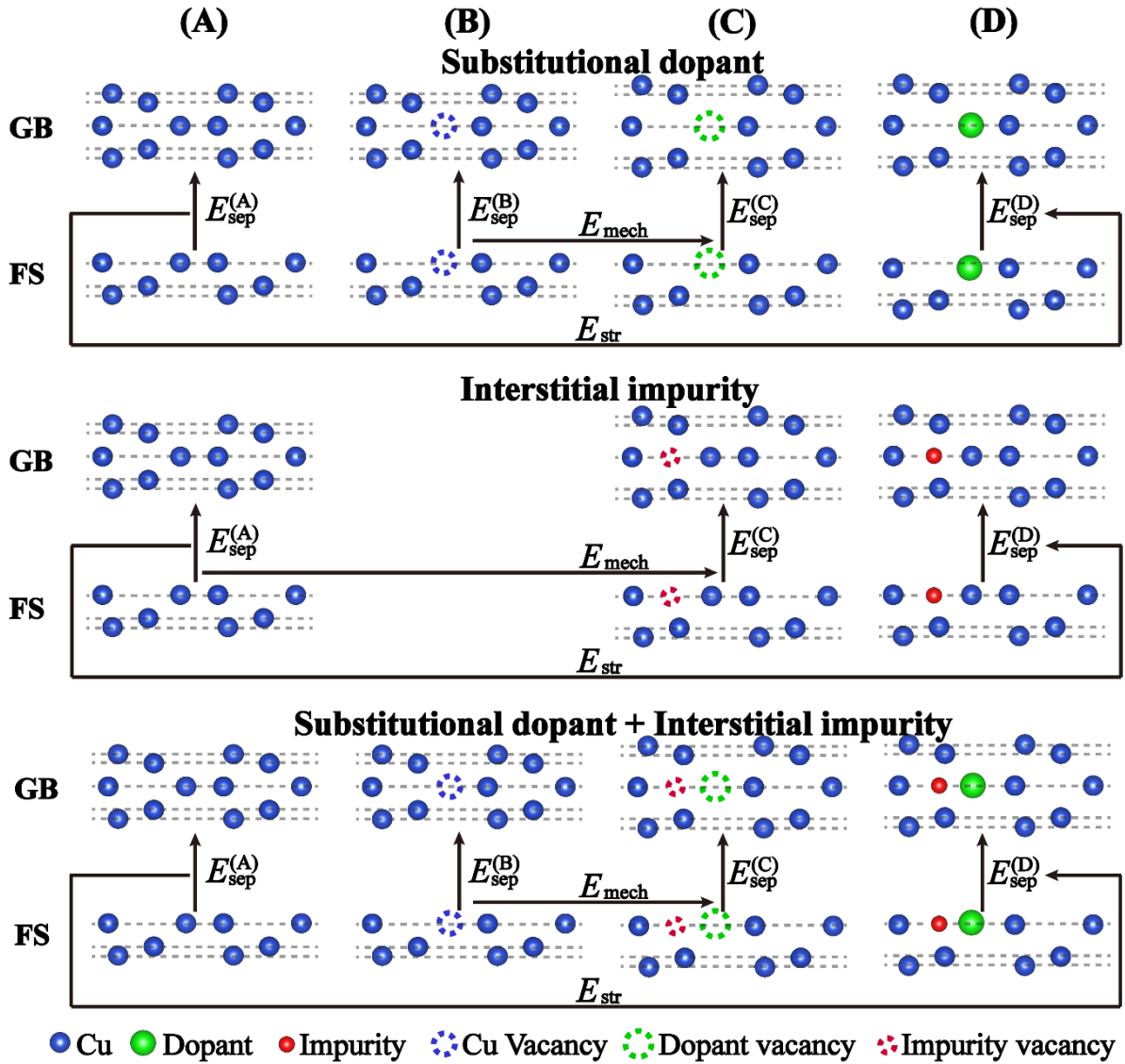

Fig. 2. Schematic illustrations for the calculations of the overall strengthening energy, as well as the mechanical and chemical contributions. The chemical contribution is what remains of the strengthening energy once the mechanical contribution has been subtracted.

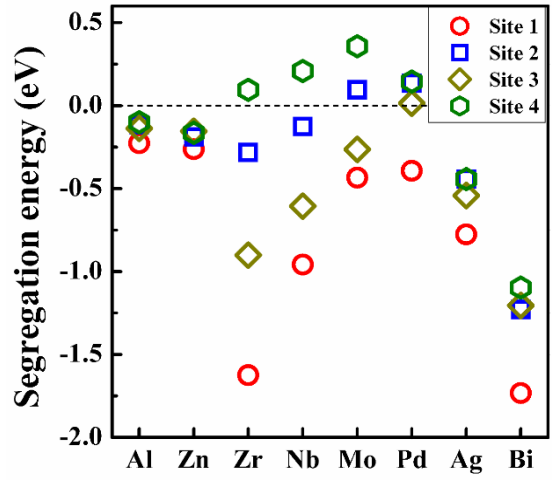

Fig. 3. The segregation energies of metallic dopants at different substitutional sites at the grain boundary.

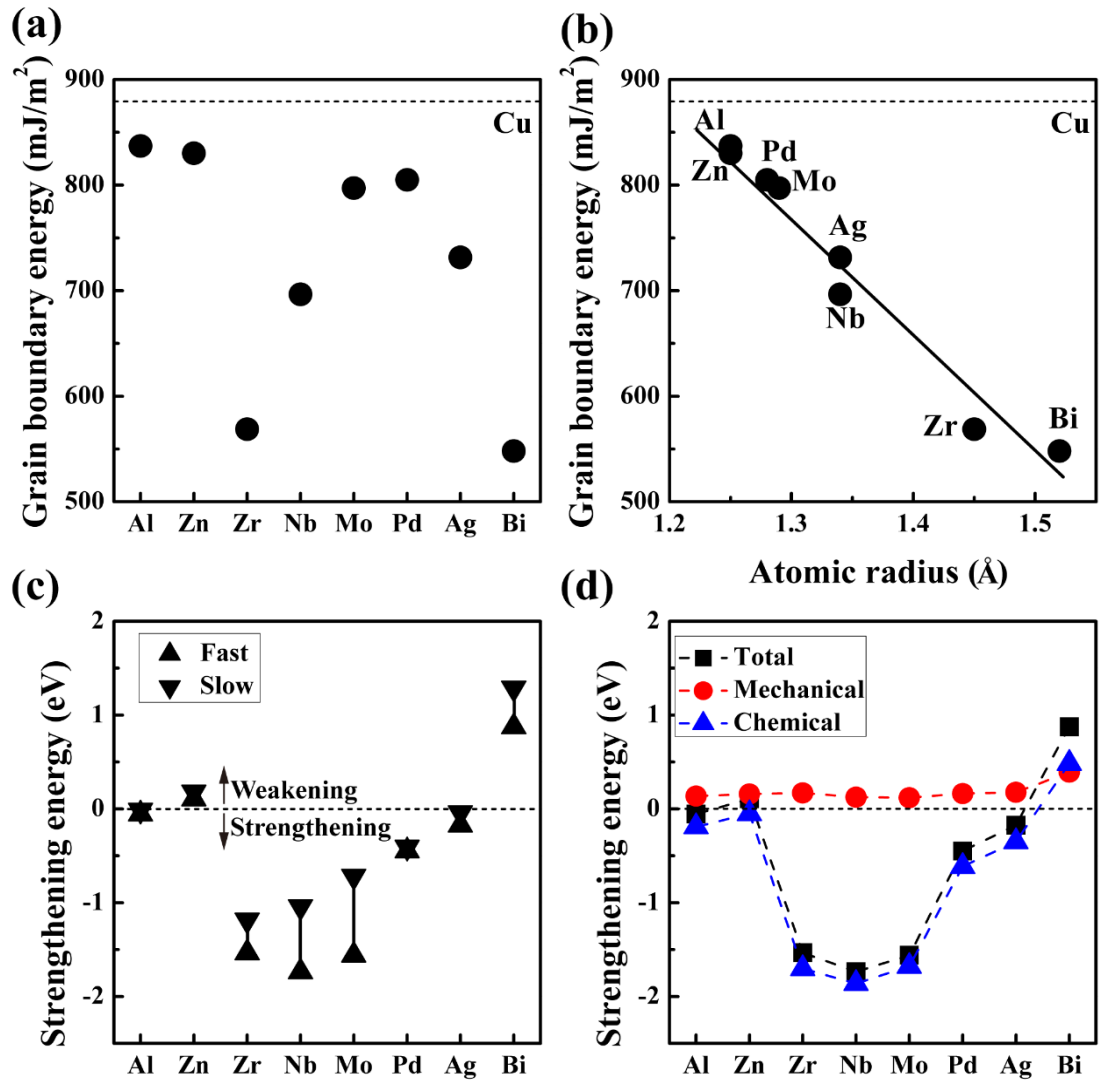

Fig. 4. (a) The grain boundary energies of the lowest-energy grain boundaries with metallic dopants. (b) The relationship between grain boundary energy and the atomic radius of dopants. (c) The strengthening energies of the metallic dopants-segregated grain boundaries under the Fast and Slow fracture progresses. (d) The mechanical and chemical contributions of the total strengthening energies.

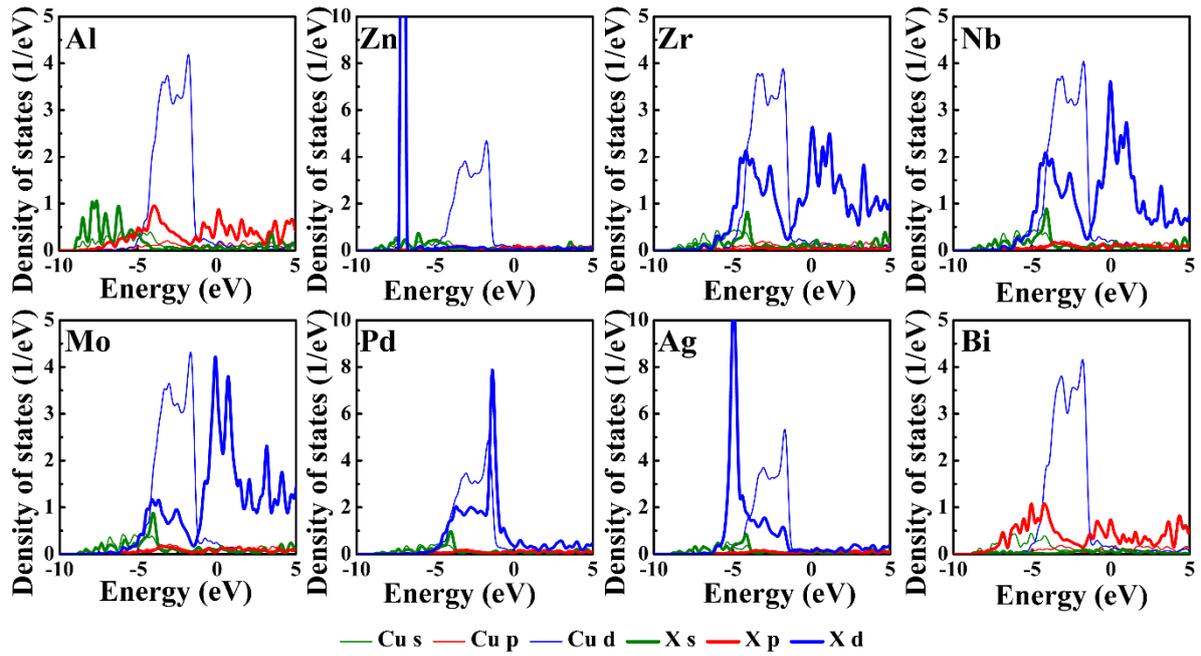

Fig. 5. The density of states for the dopant X and the closest Cu atom in site 2. The density of states data for Cu appear as thin lines, while the density of states data for each dopant X appear as thick lines.

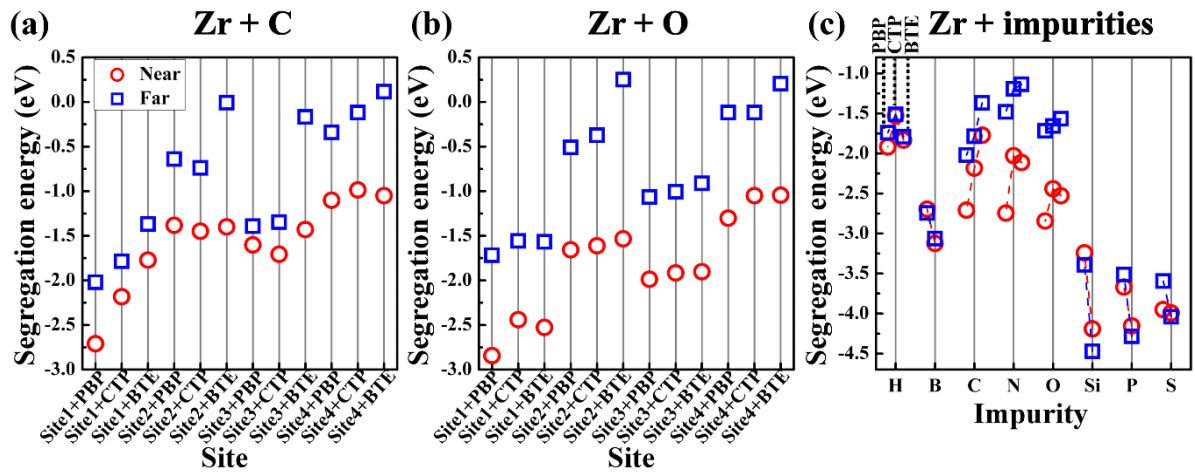

Fig. 6. The segregation energies of grain boundaries with Zr at different substitutional sites and (a) C and (b) O at different interstitial sites with the Near and Far configurations.   (c) The segregation energies of grain boundaries with Zr at site 1 and impurities at different interstitial sites with the Near and Far configurations.

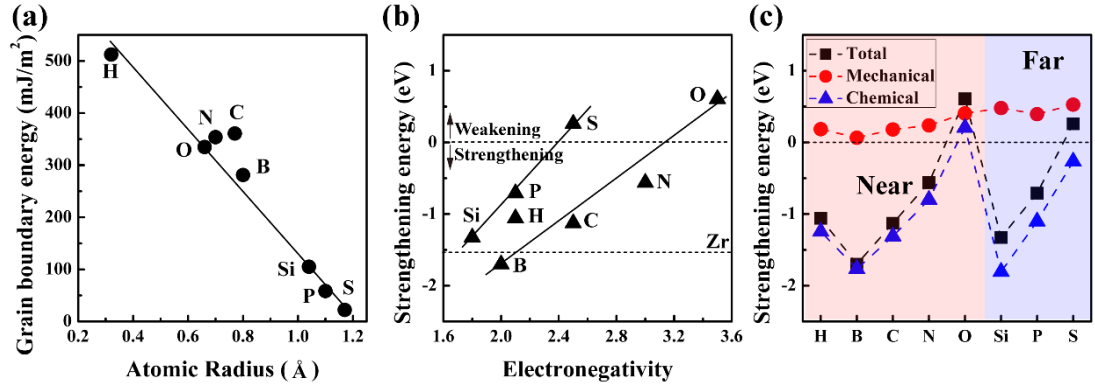

Fig. 7. Grain boundary energies and strengthening energies of the grain boundaries with Zr plus impurities. (a) The relationship between the grain boundary energy and the atomic radii of the impurities. (b) The relationship between the strengthening energy and the electronegativity of the impurities. (c) The mechanical and chemical contributions to the strengthening energies. The labels Near and Far denote the Near and Far configurations.

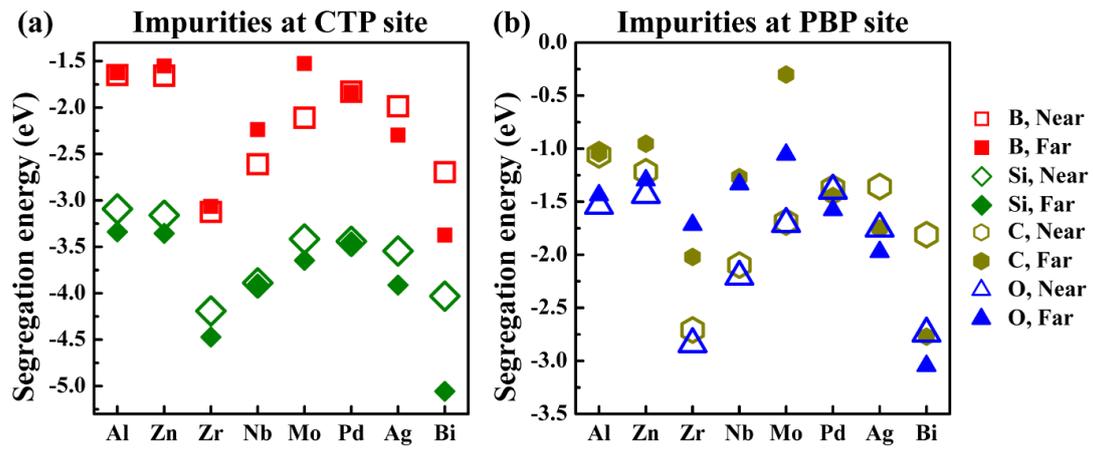

Fig. 8. The segregation energies of grain boundary with dopants and (a) impurities at the CTP sites and (b) impurities at the PBP sites with the Near and Far configurations.

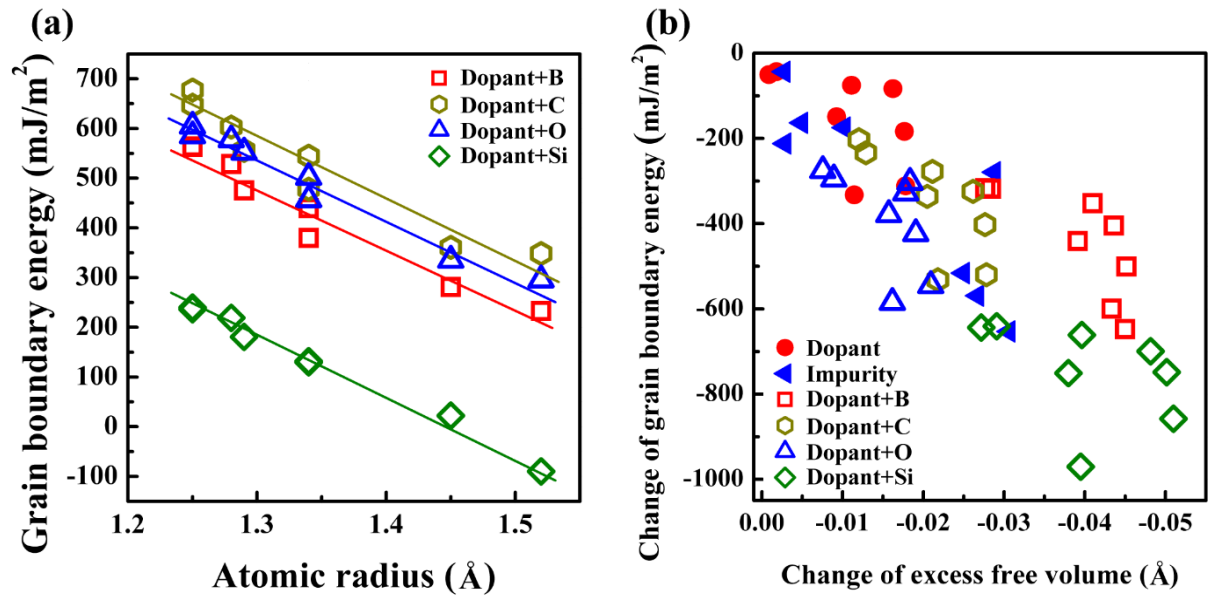

Fig. 9. (a) The relationship between the grain boundary energy and the atomic radii of dopants for the grain boundaries with dopants plus impurities. (b) The relationship of the change of grain boundary energy and the change of excess free volume.

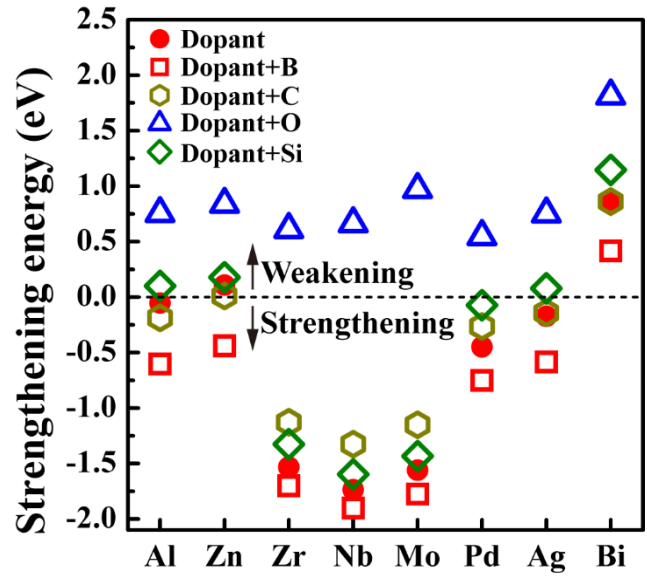

Fig. 10. The strengthening energies of grain boundaries with dopants plus impurities.

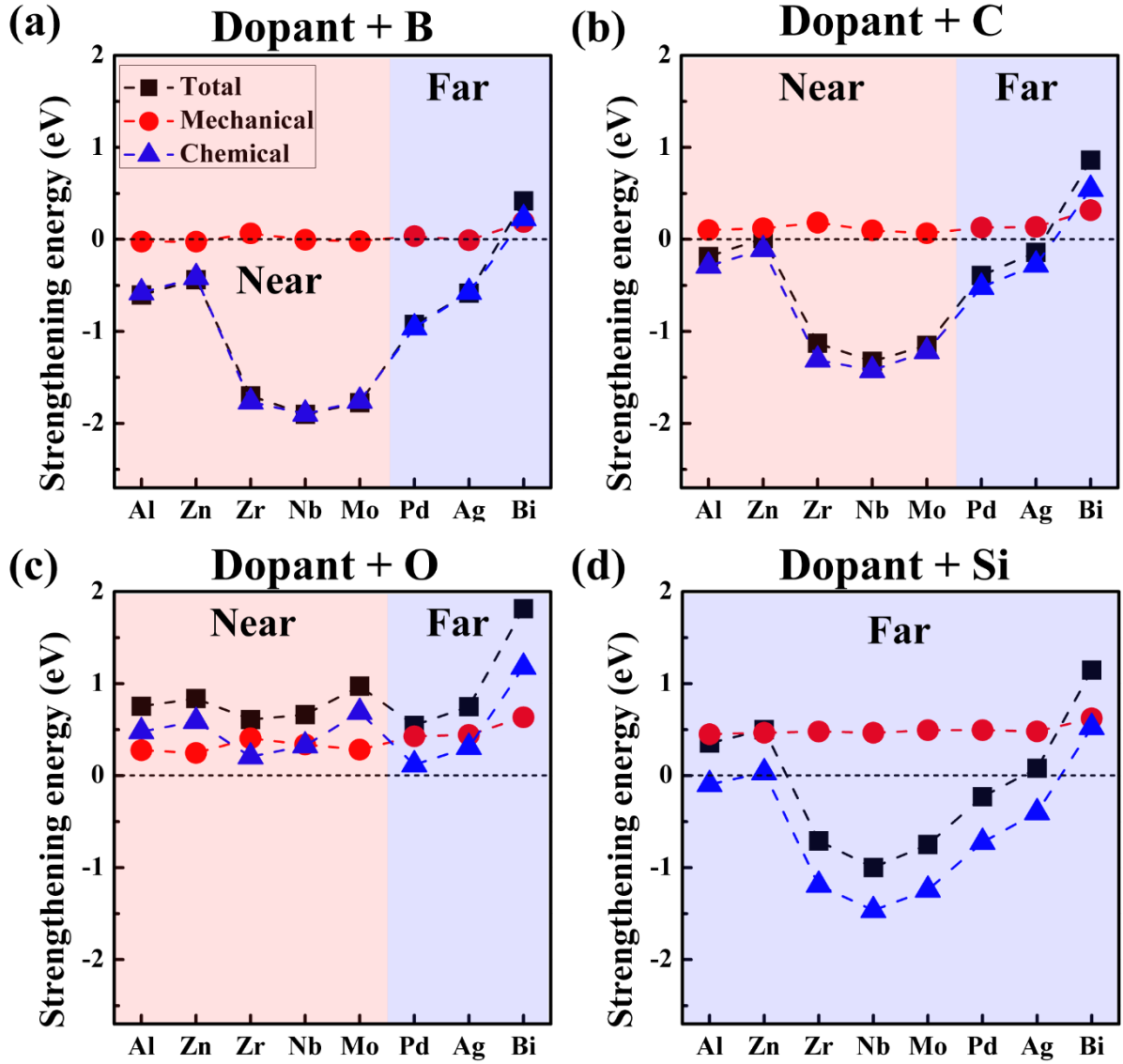

Fig. 11. The mechanical and chemical contributions to the total strengthening energies for grain boundaries with dopants plus (a) B, (b) C, (c) O, and (d) Si. The labels Near and Far denote the Near and Far configurations.

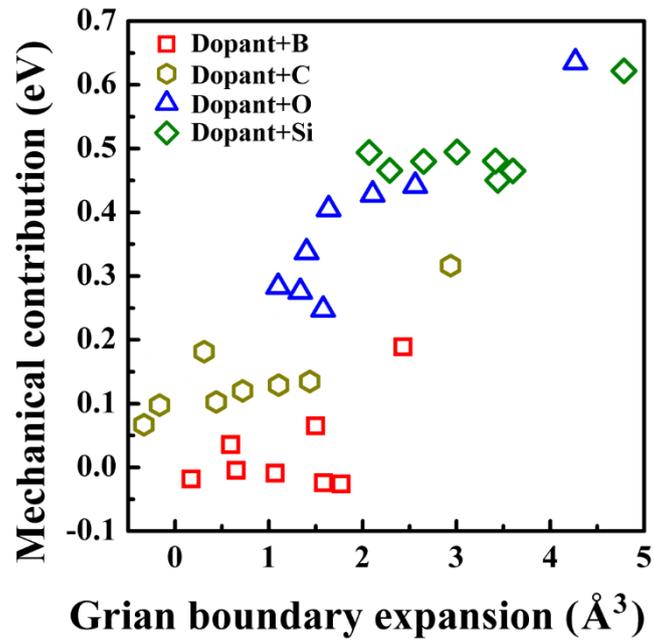

Fig. 12. The relationship between the mechanical contribution and the grain boundary expansion of grain boundaries with dopants plus impurities.

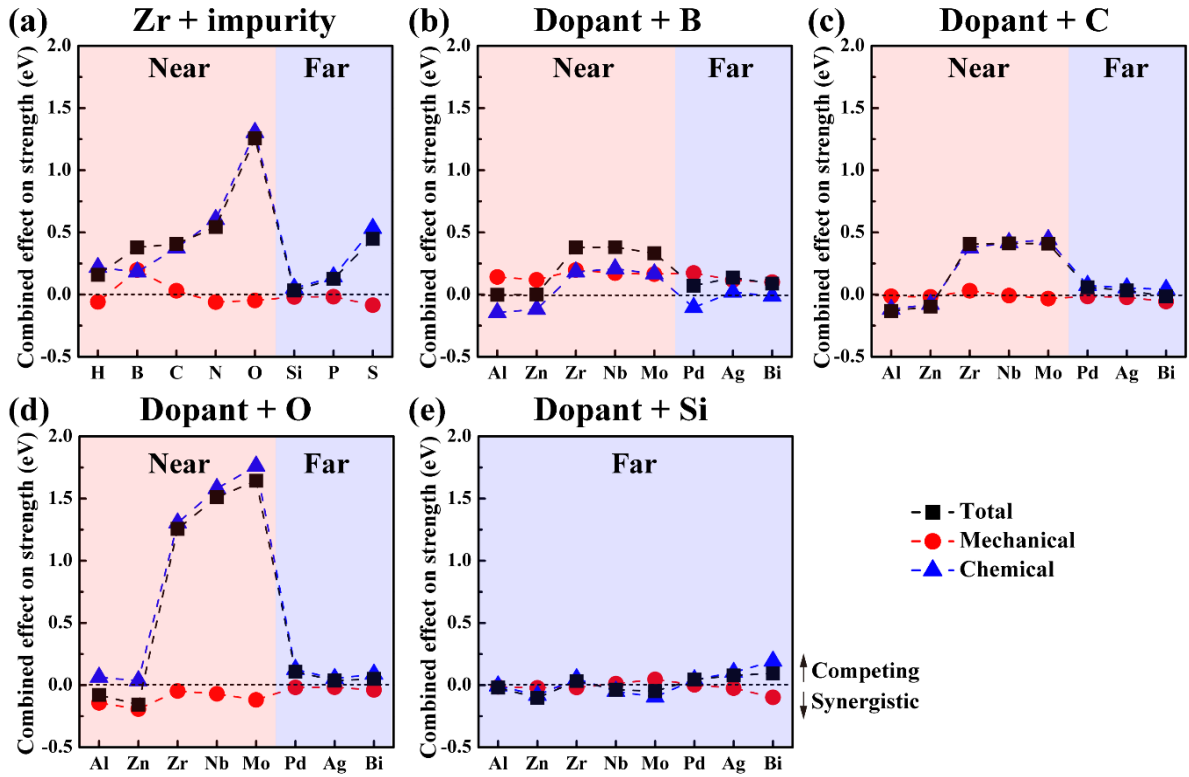

Fig. 13. The combined effects on grain boundary strength of (a) Zr and impurities, (b) dopants and B, (c) dopants and C, (d) dopants and O, and (e) dopants and Si. The labels Near and Far denote the Near and Far configurations.

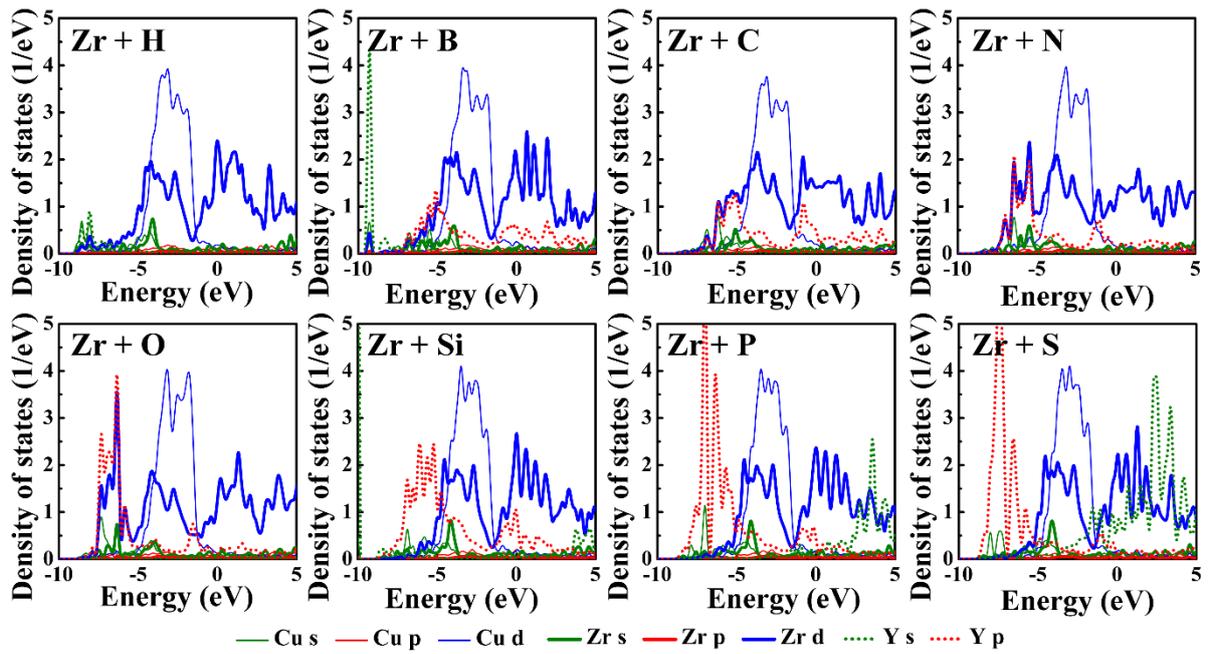

Fig. 14. The density of states data of the Zr atom, the impurity Y, and the closest Cu atom in site 2. The density of states for Cu, Zr, and the impurity Y appear as thin curves, thick curves, and dashed curves, respectively.